\documentclass[aps,preprint,eqsecnum,tightenlines,floatfix]{revtex4}
\usepackage{graphicx}

\def\ls{{\bf L}\cdot{\bf S}}

\def\beq{\begin{equation}}
\def\eeq{\end{equation}}
\def\ba{\begin{eqnarray}}
\def\ea{\end{eqnarray}}
\def\vij{v_{ij}}

\def\rij{r_{ij}}
\def\opij{O^p_{ij}}
\def\vpij{v_p(r_{ij})}

\def\vet{v18}
\def\vep{v8^{\prime}}
\def\vsp{v6}
\newcommand{\boldsigma}{\mbox{\boldmath$\sigma$}}

\newcommand{\boldnabla}{\mbox{\boldmath$\nabla$}}

\begin{document}

{

\title{Quantum Monte Carlo Calculations of Neutron Matter}
\author {J. Carlson} 
\affiliation {Theoretical Division, \\ 
  Los Alamos National Laboratory, 
  Los Alamos, New Mexico 87545} 
\author {J. Morales, Jr, V. R. Pandharipande and D. G. Ravenhall } 
\affiliation { Department of Physics,  
  University of Illinois at Urbana-Champaign, \\
        1110 W. Green St., Urbana, IL 61801, U.S.A.}

\date{\today}

\begin{abstract}

Uniform neutron matter is approximated by a cubical box containing 
a finite number of neutrons, with periodic boundary conditions.
We report variational and Green's function Monte Carlo calculations of the 
ground state of fourteen neutrons in a periodic box using the Argonne $\vep $ 
two-nucleon interaction at densities up to one and half times the nuclear 
matter density.  The effects of the finite box size are 
estimated using variational wave functions together with 
cluster expansion and chain summation techniques.  They are small at 
subnuclear densities.  We discuss the expansion of 
the energy of low-density neutron gas in powers of its Fermi momentum.  This 
expansion is strongly modified by the large $nn$ scattering length, and does 
not begin with the Fermi-gas kinetic energy as assumed in both   
Skyrme and relativistic mean field theories.  The leading term of neutron 
gas energy is $\sim$ half the Fermi-gas kinetic energy.  
The quantum Monte Carlo 
results are also used to 
calibrate the accuracy of variational calculations employing Fermi hypernetted 
and single operator chain summation methods to study nucleon matter 
over a larger density range, 
with more realistic Hamiltonians including three-nucleon interactions. 


\end{abstract}

\maketitle

\section{Introduction}

Since the discovery of neutron stars in 1967 there has been a continued interest 
in calculating the properties of neutron matter from realistic models of nuclear 
forces \cite{PR95,HP00}.  It is difficult to extrapolate the data on bound  
nuclei using energy-density functionals to estimate the equation of state $E(\rho)$ of 
pure neutron matter.  Various Skyrme and relativistic energy-density functionals  
which fit the binding energies and radii of nuclei available in laboratories at 
present give rather different $E(\rho)$ for neutron matter.  These energy-density 
functionals also predict 
different properties of nuclei near the neutron drip line \cite{WNX}, which may be 
synthesized in the near future using radioactive ion beams.  Theoretical predictions  
of the neutron matter $E(\rho)$ have been used to constrain 
the energy density functionals used to study neutron rich nuclei. 

The two-neutron interaction is strong and highly spin dependent. Therefore 
calculating the neutron matter $E(\rho)$ is a challenging 
many-body problem,  though in some ways it is simpler than that of symmetric nuclear 
matter.  Neutron matter properties have been calculated recently with  
Brueckner theory \cite{Engvik97,BBB97} and with variational methods using chain 
summation techniques \cite{WFF88,APR98}.  There is good agreement between the 
results of these two methods \cite{HP00}, and recent high precision models of $nn$ 
interaction give rather similar neutron matter $E(\rho)$ with the lowest order 
Brueckner method \cite{Engvik97}.  The results for symmetric nuclear matter, 
however, have more model dependence.  The $E(\rho)$ of high density neutron 
matter is also sensitive to the lesser known three neutron interaction \cite{APR98}. 

The Brueckner and variational methods use different expansions. From the results 
of two- and three-hole line contributions, the cluster expansion
in Brueckner theory appears to converge \cite{SBGL98}, though contributions of 
clusters with more than three neutrons have not been calculated.  In contrast, 
the cluster expansion of the energy expectation value of neutron matter 
used in the variational method has a rather poor 
convergence when the optimum variational wave function, $\Psi_V$ is used. In 
the latest calculations \cite{MPR02} two- and three-body cluster contributions are 
calculated accurately, while those of $\geq 4$-body clusters are summed approximately 
with hypernetted and single-operator 
chain summation methods.  The convergence rate of the expansion 
is sensitive to the range of correlations in $\Psi_V$.  Hence it is often possible to 
use shorter range correlations, which give a more convergent cluster expansion 
together with a variational energy within a few \% of the optimum minimum.  Thus, even 
though the results of these two methods are in agreement within several \%, the 
theoretical error in the treatment of long range correlations is not well estimated. 

In the past few years it has been possible to calculate the energies of all the 
bound states of nuclei having 
up to ten nucleons with errors estimated to be $\alt$ 2 \% using the Green's 
function Monte Carlo (GFMC) method \cite{WPCP00,PWV02}.  Results of these 
calculations are being used to construct realistic models of three-nucleon 
interactions \cite{PPWC01}.  The computational effort necessary for a nuclear 
GFMC calculation scales approximately with $2^A A!/(N!Z!)$ for a system with 
$N$ neutrons, $Z$ protons and $A=N+Z$.  The factor $2^A$ comes from the number of 
spin states of $A$ nucleons and $A!/(N!Z!)$ is the number 
of charge conserving isospin states. 
In the present work we report calculations of the ground state of 14 neutrons 
in a periodic box (PB) with the GFMC method considering all the $2^{14}$ spin states. 
We have also used the auxiliary field diffusion Monte Carlo (AFDMC) method 
proposed by Schmidt and Fantoni \cite{SF99} to calculate the ground 
state energy.  In this method one effectively samples the $2^{14}$ spin states 
stochastically.  The computational effort of AFDMC scales with $A^3$, and thus it 
can be used to study systems with larger values of $A$ \cite{FSS01}.  The present 
AFDMC calculations seem to have larger errors than the GFMC; however, it may be 
possible to improve their accuracy. 

The interactions and the variational wave functions used in this work are described 
in sect. II. The quantum Monte Carlo calculations, variational (VMC), GFMC and AFDMC, are 
described briefly in section III, 
where we present results at $\rho=0.04,~0.08,~0.16$ and 0.24 fm$^{-3}$. 
The details of these methods have been presented previously; here we
simply describe the additional techniques used to calculate results for 14 neutrons
in a PB, and discuss several tests of the calculations.  
The total energy and the potential energy expectation values 
are reported for each density.

The variational calculations using chain summation methods (VCS) are reported 
in section IV.  In this section we also discuss the difference between the 
density matrices of 14 neutrons in a PB and of uniform gas (UG) with large 
number of neutrons.  The smallness of this difference makes the 14-particle 
PB a useful approximation to UG.  The difference between the energy per neutron 
in the PB and in UG is estimated using variational wave functions.  It is small 
at subnuclear densities, $\rho \leq \rho_0 = 0.16$ fm$^{-3}$, 
but significant at 1.5 $\rho_0$.  
The comparison of VCS results with the QMC suggests that the former can have errors 
up to $\sim$ 10 \%.  

The pair distribution functions obtained from VMC and GFMC calculations are 
compared in section V.  These indicate that neutron matter has strong correlations 
even at small densities, as expected from the large scattering length, $a \sim -18$ fm in the 
$^1S_0$ state.  The results for the $E(\rho)$, extrapolated to the UG 
limit, are presented in section VI.  Here we also discuss the expansion of the $E(\rho)$ 
in powers of $k_F$.  When $|ak_F| < 1$, this expansion begins with the  
Fermi-gas kinetic energy, $T_{FG}=0.3 k_F^2/m$.  However, at densities of interest in 
nuclear or neutron star physics $|ak_F| >> 1$, and the expansion of the $E(\rho)$ 
seems to begin with $\sim T_{FG}/2$, which is the estimated UG 
energy for a short range interaction with scattering length 
$(-a) \rightarrow \infty$.  This approximation to nuclear forces in low-density 
neutron-gas was suggested by Bertsch \cite{Bertsch}. 

The accuracy of the present calculation is discussed in the last section VII.  At 
densities $\leq \rho_0$ the GFMC calculation appears to be well converged and 
presumably has an accuracy of $\sim 2$ \% for the energy of normal neutron 
matter.  However, there is exceptionally strong pairing in dilute Fermi gases with 
$ak_F \rightarrow -\infty$ \cite{ERS97}, and their superfluid state can have 
energies below those of the normal state by $\sim$ 10 \% \cite{dfg}.  The main error in the 
predicted energy of low density neutron matter, where three-body forces are small, 
is likely to be due to the neglect of the superfluidity in the present calculation.

\section{Interactions and Variational Wave Functions}

We have used the Argonne $\vep$ two-nucleon interaction \cite{PPCPW97} 
in this work.  This simplified interaction 
equals the isoscalar part of the realistic Argonne $\vet$ interaction in all the 
S- and P-waves as well as in $^3D_1$ and its coupling to $^3S_1$.  In neutron 
matter this interaction can be written as an operator with four terms:
\ba
\vij &=& \sum_{p=1,4} \vpij \opij ~, \\
O^{p=c,\sigma,t,b}_{ij} &=& 1,~\boldsigma_i \cdot \boldsigma_j,~S_{ij},~{\bf L} 
		    \cdot {\bf S}~. 
\label{e:op4}
\ea
Here $S_{ij}$ and ${\bf L}\cdot {\bf S}$ are the usual tensor and spin-orbit 
operators.  In the calculations using the PB boundary condition, 
the interaction is truncated at 
$\rij=L/2$, where $L$ is the length of the cubic box holding 14 
neutrons $(L^3\rho=14)$ : 
\beq
v(\rij) = v(\rij)~\theta(\frac{L}{2}-\rij) + v(\rij)~\theta(\rij-\frac{L}{2})~. 
\eeq
The contribution of the long range part, $v(\rij)~\theta(\rij-\frac{L}{2})$, to 
the $E(\rho)$ of UG is estimated using variational calculations.  While at low densities 
this contribution is small, at $\rho = 0.24$ fm$^{-3}$ it becomes comparable 
to the total $E(\rho)$ in magnitude.  

The variational wave function $\Psi_V$ used in this work has the form:
\beq
\Psi_V = \left( {{\cal S}} \prod_{i<j} F_{ij} \right) \Phi~, 
\label{eq.psiv}
\eeq
where $\Phi$ is the noninteracting Fermion wave function.  In UG calculations 
$\Phi = \Phi_{FG}$, the Fermi-gas wave function, while in calculations 
using the PB boundary conditions $\Phi = \Phi_{PB}$.  It 
has 14 neutrons occupying spin up and down states with momenta
\beq
{\bf k} = 0,~ \pm k_B \hat{\bf x},~\pm k_B \hat{\bf y},~\pm k_B \hat{\bf z}~.
\eeq
Here $k_B=2\pi/L$ and $\hat{\bf x},\hat{\bf y}$ and $\hat{\bf z}$ are unit vectors. 

The ${{\cal S}}\prod $ denotes a symmetrized product of the 
noncommuting $F_{ij}$ pair correlation operators. 
In VCS calculations they have four terms involving the four operators of 
Eq. (\ref{e:op4}): 
\beq
F_{ij} = \sum_{p=c,\sigma,t,b} \beta_p~f_p(\rij) \opij ~. 
\eeq
The correlation functions $f_p(\rij)$ are obtained by solving two-body Schr\"{o}dinger-like 
equations \cite{WFF88}, and have three parameters, $d, d_t$ and $\alpha$.  
They correspond to the range of all but the tensor correlations $(d)$, 
range of tensor correlations $(d_t)$, and the average quenching of 
spin-dependent interactions in matter $(\alpha)$.  

In the case of UG, the  
values of $d,~d_t$ and $\alpha$ are determined by minimizing the energy 
with the VCS method for $\beta_p=1$.  Constraints imposing conservation of 
nucleons are used during this minimization \cite{WFF88} to prevent the 
$F_{ij}$ from going into regions where the chain summation approximation 
fails.  The parameters $\beta_{\sigma},~\beta_t$ and $\beta_b$ provide additional 
variation of $F$; $\beta_c$ is not a variational parameter since $F(r_{ij}\rightarrow \infty)=1$. 
The $\beta_{p \neq c}$ parameters were not used in recent 
calculations \cite{APR98} since they do not lower the energy significantly after 
optimizing $d,~d_t$ and $\alpha$.  They are used here for the following reason. 
The optimum values of $d$ and $d_t$ in UG are $> L/2$. However, 
in VMC as well as VCS calculations using the PB 
boundary condition the $d$ and $d_t$ must be $\leq L/2$. 
In all PB calculations we use $d=d_t=L/2$, and vary the $\alpha$ and $\beta_{p \neq c}$ 
to minimize the energy. 

In VMC calculations the spin-orbit correlations in the $F_{ij}$ are neglected 
due to computational 
difficulties associated with the gradient operator in ${\bf L}$.  These calculations 
use the $\vsp$ interaction obtained by dropping the spin-orbit term 
in the $\vep$.  There results are compared with those of VCS with the same $F$ to 
test the accuracy of the chain summation approximation.  
The complete $\vep$ interaction is used in the 
GFMC calculations where the spin-orbit correlations are 
generated by propagation in imaginary time as discussed in the next section.

\section{Quantum Monte Carlo Calculations}

  Quantum Monte Carlo methods have often been used to study infinite systems
of either fermions or bosons at both zero and finite-temperature.  Examples 
include atomic liquid $^3$He and $^4$He \cite{lheqmc}, 
the electron gas, \cite{egasqmc} as well
as a myriad studies of lattice models in condensed matter theory. They have
proven remarkably successful at studying the equation of state of 
strongly-interacting systems, and have also been used to explore 
phase transitions, momentum distributions, static and dynamic response, etc.
Although studies of fermion systems are usually treated via approximate
fixed-node \cite{fixednode} or constrained path (CP) \cite{constrainedpath}  
methods, these approximations
can often be quite accurate.

  The nuclear many-body problem is more difficult than all the cases
listed above, because of the strong spin-isospin 
dependence of the interaction.  Instead of a 
single function of the 3A coordinates of the particles, the wave 
function of simple systems, 
the nuclear state is described with a set
of (complex) amplitudes dependent upon the spins and isospins of the nucleons.
This complexity has been handled successfully for few-body (A $\leq$ 10) 
nuclei \cite{PWV02} by simply summing explicitly over all these amplitudes.
Monte Carlo is then used to evaluate the 3A-dimensional spatial integrals.

   Variational Monte Carlo (VMC) calculations evaluate the energy and
other observables through the use of the Metropolis Monte Carlo method.
The method is described in detail in \cite{WPCP00}, the basic idea being 
to generate points in the 3A-dimensional configuration space distributed 
with the probability density of a weight function $W({\bf R})$.  
Here ${\bf R}$ is the 3A-dimensional configuration 
vector ${\bf r}_1,...{\bf r}_A$. The 
expectation values of operators are obtained as averages over the sampled
points ${\bf R}_i$:
\begin{equation}
\langle O \rangle = 
\frac { \sum_i \langle \Psi  ({\bf R}_i) | O | \Psi ({\bf R}_i) \rangle  / W ({\bf R}_i)}
{ \sum_i \langle \Psi ({\bf R}_i) |  \Psi ({\bf R}_i) \rangle  / W ({\bf R}_i)}.
\end{equation}
The optimum weight function in most cases is the square of the wave function 
$\langle \Psi ({\bf R})|\Psi({\bf R}) \rangle $ for which the denominator of 
$\langle O \rangle $ has zero variance.  For a system of 14 neutrons the 
$\langle \Psi ({\bf R})|\Psi({\bf R}) \rangle $ is a sum of squares of the 
$2^{14}$ spin amplitudes.

    This method grows exponentially in computational time 
with increasing A, and 
present-day computers limit practical simulations to roughly 14 neutrons.
This is somewhat larger than the largest nuclei handled to date because
there is only one isospin component to the wave function.   Another
limitation of these initial calculations is that we have dropped the
$L \cdot S$ pair correlation functions, as they depend upon the momentum
of the particles in the pair.  A complete evaluation of these terms would
be difficult because the derivative operators in one pair correlation
function can, in principle,  act on all other pair correlations.  
This limitation is not very important at low densities, but can
be quite significant at higher densities. The variational wave function
cannot adequately describe p-wave pairing of the neutrons which appears
to be important at nuclear densities and above.  It may
be possible to construct a simplified wave function which includes
most of these correlations in the future.

    Green's function Monte Carlo (GFMC) methods are then used to 
obtain the ground-state energy and other properties for the 14 neutrons
with periodic boundary conditions.  The method is the same as that
used for light nuclei \cite{WPCP00},  with only very minor modifications 
used to implement the periodic boundary conditions.   The basic idea
is to sample a wave function $\Psi (\tau)$ by evaluating path
integrals of the form:
\begin{equation}
\Psi (\tau) = \prod \exp [ - (H - E_0) \Delta \tau ] | \Psi_V>,
\end{equation}
where each step in the product evolves the system over a short imaginary time $\Delta \tau$;
after many steps $n \rightarrow \infty $, $\Psi(\tau=n \Delta \tau )$ will converge to the
true ground state of the system as long as the original (variational)
wave function is not orthogonal to it.  Because we are studying systems
at densities higher than equilibrium nuclear density $\rho_0$, 
the time step used here is 0.00025 MeV$^{-1}$, or 
1/2 the time step typically used in nuclear calculations.
Again the spatial integrals are done with Monte Carlo, using a 
sum over many configurations with different spatial coordinates 
$\bf R$. Each configuration includes amplitudes for all the
2$^{14}$ spin states which are explicitly summed in the evaluation
of matrix elements.

Some of the GFMC results reported here are obtained with the CP 
approximation \cite{constrainedpath,WPCP00}.  
Since the neutrons are fermions, they may
exchange and produce contributions to the wave function of opposite signs,
and indeed with arbitrary complex phases.  In an exact GFMC calculation,
this leads to a statistical error that grows with $\tau$.  This problem
is more severe at higher densities (or with larger numbers of particles)
since it is then easier for a pair to interchange.

    To deal with this problem, we implement a constraint on the paths
to be included in the evaluation of $\Psi(\tau)$. For a spin-isospin
independent interaction the wave function is a scalar, and one can
perform a fixed-node calculation in which configurations where
the variational wave function is zero are discarded.  
This defines a surface within which
the evolution proceeds, and eliminates the sign problem at the cost
of introducing an approximation into the calculation.  The fixed-node
method is exact when the variational wave function has 
the true nodal surface, and provides an upper bound to the
true ground state energy.  This upper bound is often quite accurate 
because we are solving for the ``optimum'' solution subject only to the
boundary condition that the wave function is zero on a predefined surface
close to the correct one.

The nuclear case is more complex, both because the trial wave function
is a set of complex amplitudes and because we cannot evaluate the full
wave function for a given set of coordinates.  We can only evaluate it
for a specific order of pair correlation operators in Eq \ref{eq.psiv},  
as a complete set would require $[A(A-1)/2]!$ terms.  These pair orders
are sampled in both the VMC and GFMC calculations.  Fortunately the
fluctuations in samples of pair orders arise from the commutators of 
correlation operators.  These involve clusters of three or more nucleons, 
and they have a small effect on the variance. For the nuclear case, we construct 
an alternative constraint based on the overlap of each configuration
$\Psi (\tau, {\bf R}_i)$ with the sampled variational wave function.
Configurations with negative overlap are discarded along with those with correspondingly 
small positive overlap, ensuring that the average overlap of the discarded 
configurations with the trial wave function is zero.  This yields a stable simulation,
and the calculation can proceed out to quite large imaginary time, much
larger than the inverse gap in the system. This approximation is not
guaranteed to produce an upper bound to the ground-state energy,
though it has proven to be quite accurate for few-body nuclei.
The results obtained using this method are labeled with CP. 

The CP approximation is tested by removing the constraint.
The configurations generated
by the CP calculation are evolved further in the imaginary time
$\tau$ without constraint.  The fermion sign problem makes this
calculation more difficult for increasing density and for increasing
$\tau$.  In principle we can evaluate the energy for a much larger
$\tau$ at low density.  In practice the low-density calculations 
appear to be well converged at fairly small imaginary time.
Of course the total unconstrained imaginary time propagation is
quite small here, typically 0.005 MeV$^{-1}$, and hence only fairly
high-energy excitations are removed by this procedure.
The results of these unconstrained (UC) GFMC calculations are the most 
accurate of the presented results.  

In this work we use $\tau$ to denote the time after the CP propagation. 
The CP GFMC propagation starts at a large negative $\tau$ and ends at 
$\tau=0$.  The propagation time of CP GFMC is large enough to ensure 
convergence; however that of UC GFMC is limited by the growth in 
statistical errors due to the fermion sign problem.

Results for the VMC and GFMC calculations at different
densities are presented graphically in Figure \ref{fig:evstau}.
The upper square point at $\tau =0$, at each density, 
is the VMC results for the $\vsp$ Hamiltonian, 
while the lower square point, also at $\tau = 0$, shows the CP GFMC
results. The CP GFMC energies are typically 5 - 10\% lower than 
the VMC for the $\vsp$ interaction.  
The $\vsp$ VMC and the PB variational chain summation (VCS) calculations
discussed in section IV use the same wave function.  

The UC GFMC results are plotted as a function
of unconstrained propagation time $\tau$ after the end of CP propagation.  
Circles and squares show results for the $\vep$ and $\vsp$. 
At all densities, the $\vsp$ calculations appear to be fairly stable and little change is
observed between the CP results and the unconstrained results
for larger imaginary time.   Table \ref{tab:qmcv6} lists the total
and potential energy per neutron for various densities.
The GFMC potential energies are approximately 15\% lower than the VMC 
results indicating that true ground state has more correlations than the 
present variational wave function.  
The difference in the VMC and the GFMC potential energies is more than twice 
that in the total energies as expected near the minimum.  
The results of VCS calculations are also listed 
in Table \ref{tab:qmcv6}; they are discussed in section IV. 

The results for $\vep$ interaction are given in Table \ref{tab:qmcv8}  
and Figure \ref{fig:evstau}.
The VMC rows in this table give results with the variational wave function 
for the $\vsp$ potential without any spin-orbit correlation.  With this 
wave function the expectation value of the spin-orbit interaction, 
$\langle v_{L \cdot S} \rangle$, is small and positive.  
It in nonzero due to the tensor correlations.  In contrast the variational 
wave function used in the VCS calculations contain spin-orbit correlations 
which give significant negative $\langle v_{L \cdot S} \rangle$.  
The $L \cdot S$ correlations absent in the $\vsp$ variational wave function 
are partly generated via the CP propagation as can be seen from the 
GFMC-CP $\langle v_{L \cdot S} \rangle$ values.  
However, the constraint imposed by the $\vsp$ wave function hinders their growth.  
After the constraint is removed, the spin-orbit correlations increase 
substantially, and we obtain significantly more attaction from the $v_{L \cdot S}$.  
This will be more evident 
in the comparison of pair distribution functions in section VI.
The UC GFMC energy decreases with $\tau$ (Fig. \ref{fig:evstau}), 
and at $\rho \geq \rho_0$ the growth in statistical error limits the 
UC calculation.  

We have also performed calculations with different input correlation
functions in the trial wave function.  Their results for $\rho=\rho_0$ are illustrated
by the two sets of square points in Figure \ref{fig:gfmcafmc}.  
The points labeled GFMC(LR) are obtained with the trial wave function 
having pair correlation functions of range L/2, while those labeled GFMC(SR) 
have much shorter range input pair correlation functions.  There appears to be 
very little dependence of the UC GFMC results upon the choice of the range of 
input two-body correlation functions;  this has been checked for the pair distribution
functions as well.

We have also implemented the auxiliary-field Diffusion Monte Carlo (AFDMC) 
method of Schmidt and Fantoni \cite{SF99}. Since this method scales
much better with A than the GFMC method discussed here, it can be
used to treat much larger systems.  At present the trial wave function
used in these calculations includes only spin-independent Jastrow
factors times a Fermi-Gas determinant:
\begin{equation}
\Psi_J = [ \prod_{i<j} f^c (r_{ij}) ] \ \Phi_{FG}.
\end{equation}
Each configuration now has 14 two-component vectors describing the
relative amplitude and phases of the spin of each neutron.  These spins
rotate in the presence of fluctuating fields which, when summed,
reproduce exactly the results of the two-nucleon interaction.
As in the GFMC calculation, a constraint is imposed requiring a positive
overlap between the configuration at any time $\tau$ and the trial
wave function.

The two sets of UC AFDMC results shown in Figure \ref{fig:gfmcafmc} 
are obtained with two different estimators of the ground-state energy.  The growth
energy (black dots) is determined from the rate of increase/decrease in the population
with imaginary time $\tau$, while the mixed estimate (red dots) is determined by
the overlap of the configurations with the Hamiltonian acting on the trial
wave function.  These two estimates should be equal within statistical
errors for small values of the time step $\Delta \tau$.

The $\Psi_J$ is a very simple trial function and hence does not provide an accurate
constraint. The CP AFDMC energies at $\tau = 0$ are higher than the CP GFMC because of
this relatively poor constraint.  As $\tau$ increases beyond the CP propagation 
region, the energy drops
and becomes compatible with the GFMC results.  The statistical errrors
are somewhat worse, though, as each configuration contains only a single
set of 14 spin vectors rather than the 2$^{14}$ amplitudes in the GFMC.
The correlations between these amplitudes reduce the fermion sign problem,
but at the cost of an exponentially increasing computational time.
The AFDMC method has been used to study much larger systems with this
simple constraint, and also to study the spin susceptibility of neutron matter.
It could be used to determine the difference between the 
infinite-particle limit and the results for 14 neutrons.  Here, though,
we use VCS methods to calculate this difference.  In addition, the QMC
results provide a test of the VCS calculations often used in studies
with more realistic Hamiltonians that include three-nucleon interactions
and relativistic corrections.

\section{Variational Chain Summation Calculations} 

In VCS calculations of UG the expectation value of $H-T_{FG}$:
\beq
E_V-T_{FG} = \frac {\langle \Phi_{FG} | [{{\cal S}}\prod_{i<j} F_{ij}] (H -T_{FG})
	    [{{\cal S}}\prod_{i<j} F_{ij}] |\Phi_{FG} \rangle }
            {\langle \Phi_{FG} | [{{\cal S}}\prod_{i<j} F_{ij}] 
	    [{{\cal S}}\prod_{i<j} F_{ij}] |\Phi_{FG} \rangle }
\label{eq:evev}
\eeq
is expanded in powers of the short range functions $(F_{ij}-1)$ \cite{PW79}. 
The $\Phi_{FG}$ is an eigenstate of the kinetic energy operator $T=- \sum_i 
\nabla_i^2 / 2m $ with the eigenvalue $T_{FG}=0.3~k_F^2/m$, hence the terms with $T$ 
operating on $\Phi_{FG}$ are not included in the expansion. 
The $n$-body cluster contribution contains all the terms of this expansion 
having $n$ neutrons.  

The leading two-body cluster contribution to the energy of UG of neutrons  
is given by: 
\ba
E(2b) &=& \frac{\rho}{2} \int d^3\rij~C[F(\rij)\left(-\frac{1}{m} \nabla^2 
	+ v(\rij)\right)F(\rij)]  \nonumber \\
      &+& \frac{\rho}{2} \int d^3\rij~C[e_{ij}~F(\rij)\left(-\frac{1}{m} \nabla^2 
	+ v(\rij)\right)F(\rij)]\ell^2(\rij)  \nonumber \\
      &-& \rho \frac{1}{m} \int d^3\rij~C[e_{ij}~F(\rij) \boldnabla F(\rij)] 
	  \cdot \ell(\rij) \boldnabla \ell(\rij) ~. 
\label{eq:2bc}
\ea
Here $C[...]$ denotes the spin independent part, called the C-part \cite{PW79} of 
the operators inside the square parenthesis, $e_{ij}$ is the spin exchange operator: 
\beq
e_{ij} = - \frac{1}{2} \left( 1 + \boldsigma_i \cdot \boldsigma_j \right)~,  
\eeq
and $\ell(r)$ is the spatial density matrix:
\beq
\ell({\bf r}) = \frac{1}{A} \sum_i e^{i {\bf k}_i \cdot {\bf r}}~, 
\eeq
normalized such that $\ell(r=0)=1$.  It is given by the Slater function:
\beq
\ell(r)= 3[sin(x)-x~cos(x)]/x^3~;~~~~x=k_Fr~. 
\label{eq:slat}
\eeq
for the $\Phi_{FG}$. 

It is relatively simple to calculate the above 2-body 
cluster contribution without approximations.  All the terms in the 3-body cluster energy 
except those containing spin-orbit correlations can now be calculated exactly 
\cite{MPR02}.  However, all the $\geq 4$-body cluster contributions as well as the 
3-body contributions from ${\bf L}\cdot{\bf S}$ correlations are estimated approximately 
using the chain summation methods. 

Results of VCS calculations of the UG are given in Table \ref{tab:vcsug}.  These are at optimum 
values of $d$ and $d_t$, which generally exceed the $L/2$ of 14 neutron PB.  In this case 
the $E_V$ obtained with $\beta_p=1$ is within $\sim$ 2 \% of the $E_V$ with optimum 
$\beta_p$.  The contributions 
of clusters are calculated following \cite{MPR02}.  The 3- and $\geq 4$-b-static 
contributions do not include 
spin-orbit interaction and correlation terms; their contributions are listed 
in row $\geq 3$-b-${\bf L} \cdot {\bf S}$.  
The values of 3-b-static contributions calculated with the chain summation approximation 
are also given in Table \ref{tab:vcsug} for comparison.  They are typically within 
10 \% of the exact values.  
The listed values of $\geq 4$-b-static contributions include the elementary 
4-body circular exchange diagram discussed in Appendix A.  It was omitted in 
previous \cite{APR98,MPR02} calculations because it is generally small in symmetric 
nuclear matter.  However, this contribution contains the factor $s^{-3}$, where 
$s = 2,4$ is the spin-isospin degeneracy factor in neutron and symmetric nuclear matter. 
It is relatively larger in neutron matter, and its values are 
listed in Table \ref{tab:vcsug}. 

Table \ref{tab:vcsug} clearly shows that the cluster expansion of the $E(\rho)$ has 
slow convergence.  At low densities this is primarily due to the large 
$nn$ scattering length.  With the $\vep$ interaction, the
the total contribution of clusters with $n \geq 4$ is $\sim$ 30 (10) \% 
of the total energy at $\rho=$ 0.04 (0.16). 

Table \ref{tab:vcsugbf} gives the results for UG variational energy for 
$d=d_t=L/2$ and optimum values of $\alpha$ and $\beta_p$.  The cluster expansion 
has better convergence for these shorter range correlations, and the 
$E_V(d=d_t=L/2)$ is above the optimum $E_V$ by only $\sim$ 0.1 MeV for 
$\rho \leq \rho_0$, while at $\rho = 1.5 \rho_0$ it is higher by 0.3 MeV. 
At small $\rho$ the $\beta_t$ is significantly larger than 1. 

\subsection{Variational Calculations with $\Phi_{PB}$}

These calculations use the truncated interaction, $v(r_{ij})\theta(L/2-r_{ij})$, 
and correlation ranges $d=d_t=L/2$.   The $\Phi_{PB}$ is an eigenstate of the 
kinetic energy with the eigenvalue (per neutron):
\beq
T_{PB} = \frac{1}{2m}~\frac{6}{7} \left(\frac{2\pi}{L}\right)^2  
=1.014~ T_{FG}~.
\eeq
We note that the kinetic energy, $T_{PB}$ per neutron of the 14 noninteracting 
neutrons in a periodic box is only 1.4 \% larger than that of free Fermi gas.  

As in the case of the UG we expand the expectation value of $H-T_{PB}$ as a sum over 
clusters.  The leading two-body cluster contribution is given by Eq.\ (\ref{eq:2bc}) 
with the PB density matrix:
\beq
\ell_{PB}({\bf r})=\frac{1}{7}(1+2~cos(xk_B)+2~cos(yk_B)+2~cos(zk_B))
\eeq
in place of the UG density matrix given by the Slater function (Eq. \ref{eq:slat}). 
The $\ell_{PB}$ depends upon the direction of ${\bf r}$, as illustrated in 
Fig.~\ref{fig:boxl}.   
The $\ell_{PB}$ is largest for ${\bf r}$ parallel to the box 
side and smallest along the diagonal.  

The C-parts of the operators in $E(2b)$ (Eq. \ref{eq:2bc}) are spherically symmetric 
functions of $r_{ij}$, therefore the $E(2b)$ depends only on the angle averaged value of 
$\ell^2$:
\beq
\overline{\ell^2_{PB}(r)} = \frac{1}{4\pi} \int sin\theta~d\theta~d\phi~\ell^2_{PB}({\bf r})~,
\eeq
at $r \leq L/2$. The $\overline{\ell^2_{PB}(r)}$ is fairly close to $\ell^2(r)$ in UG as shown in 
Fig.\ \ref{fig:boxl}. 
Therefore the 2-body cluster contributions obtained with the UG and PB density matrices 
are not very different.  In the variational calculations with $\Phi_{PB}$ we approximate 
the contributions of all $n \geq 3$-body clusters by their values in the UG. 
The 3-b-static cluster is calculated exactly, and the rest with chain summation approximations.  
We note that Fantoni and 
Schmidt \cite{PBFHNC} have developed chain summation methods for calculations with $\Phi_{PB}$ 
sans tensor correlations.  They retain the $\ell_{PB}$ in all the many-body 
cluster contributions calculated with chain summation methods. 

The results of calculations with the $\vsp$ and $\vep$ interactions 
are given in Tables \ref{tab:v6box} and \ref{tab:v8box}.  The values of $T_{FG}-T_{PB}$ and 
\beq
\Delta E(2b) = E_{UG}(2b)-E_{PB}(2b)
\eeq
are also listed in Table \ref{tab:v8box}.  The smallness of these differences makes the 14-neutron 
periodic box a good approximation for studying uniform gas $E(\rho)$.  The variational 
parameters $\alpha$ and $\beta_p$ have essentially the same values in PB calculations 
as in UG with $d=d_t=L/2$.  The main difference 
between the UG and PB energies comes from the contribution of the long range interaction, 
$v(r_{ij})\theta(r_{ij}-L/2)$ omitted in the PB.  Its contribution denoted by $\langle 
v(r_{ij} > L/2) \rangle$ is estimated from UG calculations and listed in Table \ref{tab:v8box}.
It becomes comparable to the total $E_{UG}(\rho)$ at $\rho \sim 1.5~\rho_0$.  

\subsection{Comparison with QMC Calculations }

The results of the approximate VCS calculations are compared with those of 
QMC calculations in Tables \ref{tab:qmcv6} and \ref{tab:qmcv8} for the $\vsp$ and $\vep$ 
interactions.  Since the VMC and VCS calculations use the same wave function for the 
$\vsp$ Hamiltonian, they should ideally give the same results.  The difference between 
them is due only to the approximations in the VCS calculations.  Relative to the VMC 
results, the VCS total energy is higher by 8 \% at $\rho_0/4$ and lower by 2 \% at 
$1.5 \rho_0$.  The 
difference between the VCS and VMC potential energy, $\langle \vsp \rangle$ is similar.  The clusters 
with $\geq 3$-neutrons give a relatively larger contribution at smaller densities 
(Table \ref{tab:v6box}) due to the large $nn$ scattering length.  It is thus not surprising 
that VCS has larger errors in that region.  The GFMC-UC energies are below those of 
the VCS by 12 to 4 \% in this density range. 
We note that the differences between the GFMC-UC and VCS or VMC potential energies are 
much larger, of order 20 \%.  This is because the present variational wave functions 
underestimate the correlations in matter, as is further elaborated in the next section.  

In the case of the $\vep$ Hamiltonian, the VMC calculations do not include the 
spin-orbit correlations while the VCS do, therefore the VMC and VCS results are not comparable in 
this case. However, we can compare the GFMC-UC and VCS results.  As for $\vsp$, the 
GFMC-UC energies are lower by 10 \% at lower densities; however, at $1.5 \rho_0$ the 
VCS energy is lower  by $2 \pm 0.5$ MeV.  Most of this difference seems to come from the 
spin-orbit interaction.  The $\langle v_{ {\bf L} \cdot {\bf S}} \rangle = - 8 \pm 1$ 
and $-12$ MeV in these two calculations at $1.5 \rho_0$.  When the density 
dependence of spin-orbit correlations is neglected, the leading two-neutron cluster gives a 
contribution proportional to $\rho^{5/3}$ to $\langle v_{ {\bf L} \cdot {\bf S}} \rangle$. 
This comes about because the spin-orbit 
contributions are proportional to $\rho^{2/3}$ via the $k^2$ momentum  dependence of 
$v_{ {\bf L} \cdot {\bf S}} {\bf L} \cdot {\bf S} f_{ {\bf L} \cdot {\bf S}} {\bf L} \cdot {\bf S}$, 
and the summation over particles gives an additional factor of $\rho$ for 2-body clusters.  
The VCS results for  $\langle v_{ {\bf L} \cdot {\bf S}} \rangle$ approximately 
follow this density dependence as shown in Table \ref{tab:qmcv8}.  Up to $\rho_0$ 
the GFMC-UC   $\langle v_{ {\bf L} \cdot {\bf S}} \rangle$ also has a similar 
density dependence.  However, at $1.5\rho_0$ the GFMC-UC is smaller in 
magnitude.  It could be that higher order cluster terms become more important 
at this density and that VCS overestimates the $v_{\ls}$ contribution, or that,  
if GFMC-UC is propagated further in the imaginary time $\tau$ after the CP, 
the $\langle v_{ {\bf L} \cdot {\bf S}} \rangle$ will decrease and the GFMC-UC 
energy will go down.  In the present calculation we can not test this possibility 
because of the increase in the statistical errors due to the Fermion sign problem.  

\section{Pair Distribution Functions}

The pair distribution functions obtained from QMC calculations with the
$\vep$ 
interaction are shown in Figures \ref{fig:pairdisv8-.04} 
to \ref{fig:pairdisv8-.24}.  In each figure the circles show the results of 
VMC calculations with a wave function containing $f6$ correlations
(without
L$\cdot$S correlations).  Squares and triangles represent the result of the
constrained path (GFMC-CP), and the unconstrained (GFMC-UC).
The pair distribution functions $g_p (r)$ are given by the expectation value:
\begin{equation}
g_p (r) = N \sum_{i<j} \langle \Psi | \delta (r_{ij} - r) O^p_{ij} | \Psi
\rangle,
\end{equation}
with a normalization $N$ such that $g_1 (r) \equiv g_c(r)$ goes to one at large
distances.  The $g_{2-4}$ are denoted by $g_{\sigma},g_t$ and $g_{LS}$ 
for clarity.  The $g_c(r)$ gives the probability to find a neutron at a 
distance $r$ from another neutron since $O^{p=1}_{ij}=1$.  
In contrast $O^{p=2}_{ij}=\boldsigma_i \cdot \boldsigma_j$, thus $g_{\sigma}$ 
is proportional to the expectation value of $\delta(r_{ij}-r)
\boldsigma_i \cdot \boldsigma_j$.  Using the projection operators:
\ba
P_{S=0}&=& \frac{1}{4}(1-\boldsigma_i \cdot \boldsigma_j)~, \\
P_{S=1}&=& \frac{1}{4}(3+\boldsigma_i \cdot \boldsigma_j)~, 
\ea
the pair distribution functions in spin $S=0$ and 1 pairs are found to be:
\ba
g_{S=0}(r)&=& \frac{1}{4}(g_c(r)-g_{\sigma}(r))~, \\ 
g_{S=1}(r)&=& \frac{1}{4}(3 g_c(r)+g_{\sigma}(r))~.  
\ea
Since $g_{S=1}(r \rightarrow 0) \rightarrow 0$, the $g_{\sigma}(r)=-3g_c(r)$ at 
small $r$.  In noninteracting Fermi gases: 
\ba
g^{FG}_c(r)&=&1-\frac{1}{2}\ell^2(r)~, \\
g^{FG}_{\sigma}(r)&=&-\frac{3}{2}\ell^2(r)~, \\
g^{FG}_t(r)&=&g^{FG}_{LS}=0~. 
\ea
The VMC calculations do not have spin-orbit correlations, and give 
$g_{LS}\sim 0$. 

At $\rho = 0.04$ fm$^{-3}$, there is a very strong pairing
into spin 0 states as indicated by the large negative $g_{\sigma}$.  
It can be seen more clearly in  Figure \ref{fig:spinzero} which compares the 
$g_{S=0,1}(r)$ in neutron matter and Fermi gases.  The large peak of 
the $g_{S=0}$ is due to the large negative $nn$ scattering length; it 
should be relatively model independent and grow at smaller densities.  
This pairing is present in the variational calculations,
though underestimated by $\sim $ 25 \%.  The tensor and
$L \cdot S$ correlations are quite modest at these low densities.
There is little change between the constrained GFMC-CP and
unconstrained GFMC-UC, indicating reasonably good convergence within
this class of wave functions.   The tensor correlations are long-ranged,
extending nearly to the $L/2$ limit imposed by the periodic boundary
conditions. This same behavior is seen even when starting the GFMC with trial
wave functions having much shorter range correlations.

The correlations at $\rho = 0.08$ fm$^{-3}$
are fairly similar, though the spin correlation is not as large,
and the tensor and $L\cdot S$ correlations
are becoming more significant.  The $g_{LS}$ is essentially zero 
in the variational calculation, and underestimated in CP GFMC. 

At the largest densities considered, $\rho$ = 0.16 and 0.24 fm$^{-3}$,
the differences between the variational, GFMC-CP, and
the GFMC-UC results are quite large. Both the tensor and $L \cdot S$
correlations are quite important and significantly underestimated
in the VMC and GFMC-CP calculations.  
We see a transition from low densities, where the S-wave
interaction and spin zero pairing is dominant, to these higher
densities, where the P-wave interactions are crucial.
It could be associated with the $^3$P$_2$-$^3$F$_2$ pairing 
\cite{BCLL92} expected at higher densities.  
In VCS calculations with three nucleon interaction 
(see Fig.9 of Ref.\cite{AP97}) such a behavior is 
associated with the onset of pion condensation.  

\section{Density Dependence of Neutron Matter Energy} 

The total energy of neutron matter interacting with the $\vep$ potential is 
reported in Table \ref{tab:ldnm}.  It is obtained by adding the box
corrections 
listed in Table \ref{tab:v8box} to the GFMC-UC energies listed in Table 
\ref{tab:qmcv8}.  The ratio of neutron matter $E(\rho)$ to the
noninteracting 
neutron Fermi gas energy is also listed in Table \ref{tab:ldnm}.  This
ratio 
approaches $\sim 0.5$ at low densities. 

The properties of low density neutron matter are dominated by the large
negative 
$nn$ scattering length.  When $|ak_F| << 1$ we have the
well known low density expansion \cite{FWbook}:
\beq
E(\rho)=E_{FG}(\rho)\left[ 1 + \frac{10}{9\pi} ak_F +
\frac{4}{21\pi^2}(11-2ln2)(ak_F)^2 
+ ...\right] ~. 
\eeq
Such an expansion is not useful for neutron matter because even at
densities as low 
as 1\% of $\rho_0$, $|ak_F| > 6$.  The limit $ak_F \rightarrow - \infty$
is perhaps 
more applicable to neutron gas than the low density expansion, as
suggested by 
Bertsch \cite{Bertsch}.  In this limit it is known that:
\beq
E(\rho)=E_{FG}(\rho) \xi ~.  
\label{eq:xiefg}
\eeq 
The estimates of $\xi$ range from 0.326 \cite{baker,hh} to 0.568
\cite{baker} to 0.59 
\cite{ERS97}.  Recent quantum Monte Carlo calculations \cite{dfg} give
$\xi \sim 0.54 $ for the 
normal phase and $\xi = 0.44 \pm 0.02$ for the superfluid phase.  Most
many-body 
calculations, both Brueckner and variational, give $\xi \sim 0.5$ for 
normal neutron matter.  As an example, 
we compare the energies of neutron matter calculated with the CSM in
1981 \cite{FP81} 
with the $E_{FG}$ and results of present calculation in Figure
\ref{fig:ldnm}. 

Eq.~\ref{eq:xiefg} implies that at low densities the interaction energy 
of neutron matter becomes proportional to $k_F^2$ as is the FG kinetic 
energy.  This interaction energy is proportional to density $(=k_F^3/3\pi^2)$ 
times the volume integral 
of the effective $G$-interaction, related to the bare $v$-interaction 
by the well-known Brueckner equation $G \phi = v \psi $.  Here $\phi$ and 
$\psi$ are the unperturbed and perturbed two-nucleon wave functions.  At 
small relative momenta of interest at low-densities, $\phi=1$, and in 
vacuum $\psi = 1 -a/r$ beyond the the range $R_v$ of $v$.  When $-a/R_v >> 1$, 
as is the case for neutrons, we can neglect the $1$ in comparison and 
approximate the $\psi$ by $-a/R_v$.  The effective interaction in vacuum 
is essentially enhanced by a factor $-a/R_v$ by the large scattering length. 
In matter the effective scattering length is limited by the interparticle 
spacing of order $1/k_F$.  Thus, when $-ak_F >> 1$, the $G$ is enhanced by 
a factor proportional to $1/k_FR_v$, its integral becomes proportional 
to $1/k_F$, and the interaction energy proportional to $k_F^2$.  At 
higher densities we see a deviation from Eq.~\ref{eq:xiefg} in Fig. \ref{fig:ldnm}. 
It starts when $k_FR_v$ becomes of order 1 and the first (unit) term of $\psi$ 
can not be neglected.  When $R_v \rightarrow 0$, as in the challenge problem 
proposed by Bertsch, Eq. \ref{eq:xiefg} is valid at all densities when 
$a = -\infty$. 

Most of the nonrelativistic Skyrme as well as the relativistic mean field 
energy density functions commonly used to study nuclei and neutron star 
matter assume that $\xi=1$.  None of these therefore can reproduce the 
equation of state, $E_0(\rho)$ of pure neutron matter \cite{PR95} obtained 
from realistic interactions even at low densities.  The effective interaction 
used in mean field models must diverge as $\rho \rightarrow 0$ due to large 
$nn$ scattering length.  Energy density functionals containing such low-density 
divergences \cite{PRlect} are probably necessary to study nuclei near neutron 
drip line or in the inner crust of neutron stars.

\section{Superfluidity and Accuracy of the Present Calculations}

Quantum Monte Carlo calculations of 14 neutrons described above 
start from a correlated Slater trial
wave function.  This trial function is appropriate for the normal phase 
of Fermi liquids and is used here as the starting point for the CP GFMC 
and to obtain the constraints used to limit the Fermion sign 
problem in that calculations.   Its results are expected to give the 
equation of state of the normal phase. 
The long-wavelength properties of the correlated Slater wave
function do not include the expected superfluid properties of neutron
matter.   Generally, the energy of the superfluid phase is not 
significantly different from that of the normal one, since  
in most systems the pairing is 
confined near the Fermi surface, and involves only relatively
few particles.  Here, however, the magnitude of the $nn$ scattering length 
is very large compared to the interparticle spacing, the pairing
is exceptionally strong, and affects all the particles.  

In principle it may be possible for the unconstrained GFMC
calculations to relax to the lower energy superfluid phase 
provided there is sufficient overlap between the fourteen particle 
wavefunctions of these phases. 
However, the unconstrained calculations are limited to fairly
short paths, due to small unconstrained propogation time, 
and hence are unlikely to relax to the superfluid phase.  
In calculations with simple spin-independent, short-range
interactions with $a = -\infty$, Schmidt {\em et al.} \cite{dfg} find, 
upon including superfluid pairing into the trial wave
function, important effects,  including an $\approx$ 20\% reduction 
in the energy per particle mentioned in the previous section. 

One of the important features of the Monte Carlo approach pursued
here is that it can be extended to include BCS pairing into
the trial wave function.  Thus it will be possible to study 
properties of superfluid neutron matter 
with realistic models of the $nn$
interaction.  We are presently pursuing such studies.
The main remaining uncertainty in the equation of state of 
low density neutron matter is probably due to the neglected 
difference in the $E_0(\rho)$ of normal and superfluid phases. 

\section{Acknowledgments}
   We would like to thank K. E. Schmidt for valuable discussions.
The work of J.C. is supported by the U.S. Department of Energy
under contract W-7405-ENG-36, while that of J.M., V.R.P. and D.G.R.
is partly supported by the US National Science Foundation grant 
PHY00-98353.  
The QMC calculations reported here were performed at the National Energy
Research Supercomputer facility.

\appendix

\section{Elementary four-body circular exchange diagrams}

The 4-body diagrams in the expansion of $E_V-T_{FG}$ (Eq. \ref{eq:evev}), 
not included in the Fermi hypernetted chain summation, are called elementary 
\cite{PW79}.  In Fermi liquids they were first calculated by Zabolitzky 
\cite{zab81}.  The circular exchange diagram shown in Fig. \ref{fig:ele4cc} 
is the only one of first order in the expansion in powers of $(F^2_{ij}-1)$, 
among these.  It therefore has special importance as emphasized by 
Krotscheck \cite{krot79}, and was included in the study of nuclear matter 
structure functions \cite{npa473}.  The thick interaction line in this 
diagram represents:
\beq
I_{13}=F_{13}\left(v_{13}F_{13}-\frac{\hbar^2}{m}[\nabla^2 F_{13}]\right)~.
\eeq
We approximate the contribution of this diagram with:
\ba
\frac{\rho^3}{4} &\int& d^3r_{12} d^3r_{13} d^3r_{14} \ell_{12} \ell_{23} \ell_{34} \ell_{41} 
f_c^2(r_{12})f_c^2(r_{23})f_c^2(r_{34})f_c^2(r_{41}) \nonumber \\
&C& \left[ (e_{12}e_{23}e_{34}+e_{34}e_{23}e_{12}) I_{13}(F^2_{24}-1) \right]~. 
\ea
The term with $e_{34}e_{23}e_{12}$ takes into account the circular exchange in 
the other direction.  The spin-orbit interactions and correlations, and the 
spin correlations between neutron pairs 12, 23, 34 and 41 
are neglected in this approximation expected to have an accuracy of order 20 \%.


\begin{table}
\caption{Quantum Monte Carlo and VCS results for 14 neutrons in PB with
the 
$\vsp$ Hamiltonian (MeV per neutron).
Statistical errors are indicated in parenthesis.}
\vspace{0.4cm}
\begin{tabular}{llllll}
 \hline
Method & $\rho$~(fm$^{-3})$  &  ~~~~0.04   &  ~~~~0.08  &  ~~~~0.16   & 
~~~~0.24  \\
\hline
VMC &  $~~~\langle H \rangle $ &~~~~7.04(01) & ~~11.32(01) &~~21.39(01)
& ~~34.30(01) \\ 
GFMC-CP &                   & ~~~~6.72(01) & ~~10.64(01) & ~~19.80(02) &
~~31.90(02) \\
GFMC-UC &                   & ~~~~6.75(01) & ~~10.64(03) & ~~19.91(11)& 
~~32.15(08)\\
VCS     &                   & ~~~~7.6      & ~~11.9      & ~~21.2     & 
~~33.6     \\
\hline
VMC &  $~~~\langle \vsp \rangle $ & ~$-$9.92(03) & $-$16.17(05) &
$-$22.79(08) & $-$26.10(11)\\ 
GFMC-CP &                   & $-$11.75(08) & $-$18.64(09)  &
$-$28.01(08)  & $-$32.94(25)\\
GFMC-UC &                   & $-$11.36(13) & $-$18.17(36) &
$-$26.62(71)  &  $-$32.71(71)\\
VCS     &                   & ~$-$9.2     & $-$15.4    & $-22.7$     &
$-$26.6    \\
\hline
\end{tabular}
\label{tab:qmcv6}
\end{table}

\begin{table}
\caption{Quantum Monte Carlo  and VCS results for 14 neutrons in PB with
the 
$\vep$ Hamiltonian (MeV per neutron).}
\vspace{0.4cm}
\begin{tabular}{llllll}
 \hline
Method &  $\rho$~(fm$^{-3})$  &  ~~~~0.04   &  ~~~~0.08  &  ~~~~0.16  
&  ~~~~0.24  \\
\hline
VMC &  $~~~\langle H \rangle $ & ~~~~7.16(01)  & ~~11.678(07) &
~~21.82(12) & ~~35.02(01) \\ 
GFMC-CP &                   & ~~~~6.43(01) & ~~10.02(02)& ~~18.54(04) &
~~30.04(04) \\
GFMC-UC &                   & ~~~~6.32(03) &  ~~~~9.501(06) &
~~17.00(27)&  ~~28.35(50)\\
VCS     &                   & ~~~~7.0 & ~~10.3  & ~~17.4 & ~~26.3 \\ 
\hline
VMC &  $~~~\langle \vsp \rangle $ &  ~$-$9.74(03) &$-$15.72(6) &
$-$21.64(09) & $-$24.37(11) \\ 
GFMC-CP &                   & $-$11.85(09) & $-$18.34(11)  &
$-$27.72(15)& $-$32.34(23)\\
GFMC-UC &                   & $-$11.44(19) &$-$17.83(30) & $-$25.62(87) 
& $-$30.52(1.35)\\
VCS     &                   & ~$-$9.3 &$-$15.1 & $-$21.6  & $-$24.9\\
\hline
VMC &  $~~~\langle v_{{\bf L} \cdot {\bf S}} \rangle $ 
                            &  ~~~0.07(01) & ~~~0.26(01)  & ~~~0.13(01) 
&  ~~~0.21(01)\\ 
GFMC-CP &                   & ~$-$0.23(11) &~$-$1.37(03) 
&~$-$2.69(03)   & ~$-$4.08(08)\\
GFMC-UC &                   & ~$-$0.85(04) &~$-$2.59(15)& ~$-$6.24(50)  
& ~$-$7.98(98)\\
VCS     &                   & ~$-$0.88 &~$-$2.3&~$-$6.9  & $-$12.1\\
$-6.9(\rho/\rho_0)^{5/3}$ & & ~$-$0.68 &~$-$2.2& ~$-$6.9  & $-$13.6\\
\hline
\end{tabular}
\label{tab:qmcv8}
\end{table}

\begin{table}
\caption{Results of VCS calculations of the UG $E(\rho)$ (MeV per neutron) 
         with optimum $d$ and $d_t$, in MeV.  Contributions with an *
are 
	 estimated using chain summation approximation; those without are
calculated 
	 exactly. The bottom two rows give the approximate 3-b-static*, 
         for comparison with the exact 3-b-static,  
	 and 4-b-elementary* circular exchange contribution included in $\geq
4$-b-static*} 
\vspace{0.4cm}
\begin{tabular}{lrrrr}
\hline
$\rho$~(fm$^{-3}$)&  0.04   &  0.08  &  0.16   &  0.24  \\
$d$~(fm)          &  3.66   &  3.31  &  2.29   &  2.20  \\
$d_t$~(fm)        &  6.17   &  5.56  &  5.22   &  4.69  \\
$\alpha$          &  0.89   &  0.87  &  0.80   &  0.72  \\
$\beta_{\sigma}$  &  0.8~   &  0.8~  &  1.0~   &  1.0~  \\
$\beta_t$         &  0.9~   &  1.0~  &  0.9~   &  0.9~  \\
$\beta_b$         &  0.9~   &  0.8~  &  1.0~   &  1.1~  \\
\hline
$T_{FG}$          &   13.9  &   22.1 &   35.1  &   46.0 \\ 
2-b-Total         &~~$-$10.1  &~~$-$16.9 &~~$-$25.5  &~~$-$33.9 \\
3-b-static        &    4.9  & 6.9    &    3.7  &   3.5  \\
$\geq$ 4-b-static*&$-$2.3  &$-$3.1 &$-$ 1.5  &$-$ 1.1 \\
$\geq 3$-b-$\ls$*&    0.1  &    0.2 &    0.2  &$-$ 0.1 \\
\hline
Total $E(\rho)$   &    6.6  &    9.2 &   12.0  &   14.5 \\
\hline
3-b-static*      &    4.6  &  6.6   &   3.5   &  3.2   \\ 
4-b-elementary*  &$-$ 0.5  &$-$0.5  &   0.5   &  0.7   \\ 
\end{tabular}
\label{tab:vcsug}
\end{table}

\begin{table}
\caption{Results of VCS calculations of the UG $E(\rho)$ (MeV per neutron) 
         with $d=d_t=L/2$, in MeV.  Contributions with an * are 
	 estimated using chain summation approximation; those without are
calculated 
	 exactly.} 
\vspace{0.4cm}
\begin{tabular}{lrrrr}
\hline
$\rho$~(fm$^{-3}$)&  0.04   &  0.08  &  0.16   &  0.24  \\
$L/2$~(fm)        &  3.52   &  2.80  &  2.22   &  1.94  \\
$\alpha$          &  0.90   &  0.85  &  0.80   &  0.80  \\
$\beta_{\sigma}$  &  0.80   &  0.9~  &  0.9~   &  1.0~  \\
$\beta_t$         &  1.50   &  2.0~  &  1.0~   &  1.0~  \\
$\beta_b$         &  0.85   &  0.9~  &  1.1~   &  1.1~  \\
\hline
$T_{FG}$          &   13.9  &   22.1 &   35.1  &   46.0 \\ 
2-b-Total         &~~$-$ 9.7  &~~$-$14.9 &~~$-$23.0  &~~$-$29.9 \\
3-b-static        &    3.9  & 2.7    & $-$0.1  & $-$0.8 \\
$\geq$ 4-b-static*&$-$1.5  &$-$0.7 &$-$ 0.0  &$-$ 0.0 \\
$\geq 3$-b-$\ls$*&    0.0  &    0.0 &    0.2  &$-$ 0.4 \\
\hline
Total $E(\rho)$   &    6.7  &    9.2 &   12.1  &   14.8 \\
\hline
\end{tabular}
\label{tab:vcsugbf}
\end{table}

\begin{table}
\caption{Results of PB variational calculations with $\vsp$ interaction
truncated at $r=L/2$. 
The variational parameters are listed in Table \ref{tab:vcsugbf}, and the
energies are MeV 
per neutron.} 
\vspace{0.4cm}
\begin{tabular}{lrrrr}
\hline
$\rho$~(fm$^{-3}$)&  0.04   &  0.08  &  0.16   &  0.24  \\
\hline
$T_{PB}$          &   14.1 &   22.4 &   35.6  &   46.6 \\ 
2-b-Total         &~~$-$9.0  &~~$-$12.6 &~~$-$14.2  &~~$-$12.3 \\
$\geq$3-b-Total   &    2.5  &  2.0     &  $-$0.1   & $-$0.7 \\ 
\hline
Total $E(\rho)$   &    7.6  &   11.9 &   21.2  &   33.6 \\
\hline
\end{tabular}
\label{tab:v6box}
\end{table}

\begin{table}
\caption{Results of PB variational calculations with $\vep$ interaction
truncated at $r=L/2$. 
The variational parameters are listed in Table \ref{tab:vcsugbf}, and the
last three rows 
give the differences between PB and UG contributions. All energies are
in MeV per neutron.} 
\vspace{0.4cm}
\begin{tabular}{lrrrr}
\hline
$\rho$~(fm$^{-3}$)&  0.04   &  0.08  &  0.16   &  0.24  \\
\hline
$T_{PB}$          &   14.1 &   22.4 &   35.6  &   46.6 \\ 
2-b-Total         &~~$-$9.5  &~~$-$14.0 &~~$-$18.3  &~~$-$19.1 \\
$\geq$3-b-Total   &    2.4  &  1.9     &  0.1   & $-$1.2 \\ 
\hline
Total $E(\rho)$   &    7.0  &   10.3 &   17.4  &   26.3 \\
\hline
$T_{FG}-T_{PB}$   &   -0.2  & -0.3   &  -0.5   &  -0.7   \\ 
$\Delta E(2b)$    &   -0.0  & -0.1   &  -0.1   &  -0.1   \\ 
$\langle v(r_{ij}>L/2) \rangle$ &$-$0.1 &$-$0.8 & $-$4.5 & $-$10.7 \\ 
\end{tabular}
\label{tab:v8box}
\end{table}

\begin{table}
\caption{Neutron matter energy with the $\vep$ interaction in MeV per
neutron.} 
\vspace{0.4cm}
\begin{tabular}{lrrrr}
\hline
$\rho$~(fm$^{-3}$)& ~~~~ 0.04   & ~~~~ 0.08  & ~~~~ 0.16   & ~~~~ 0.24 
\\
\hline
GFMC-UC           &      6.3    &      9.5   &      17.0   &     28.4  
\\
Box Correction    &   $-$0.3    &   $-$1.1   &    $-$5.1   &  $-$11.5  
\\
\hline
Total $E(\rho)$   &      6.0    &      8.4   &      12.1   &     16.9  
\\ 
\hline
$E(\rho)/E_{FG}(\rho)$ &  0.43   &     0.38   &      0.34   &   0.37    
\\ 
\hline
\end{tabular}
\label{tab:ldnm}
\end{table}


\begin{figure}[tbh]
\includegraphics[width=6in]{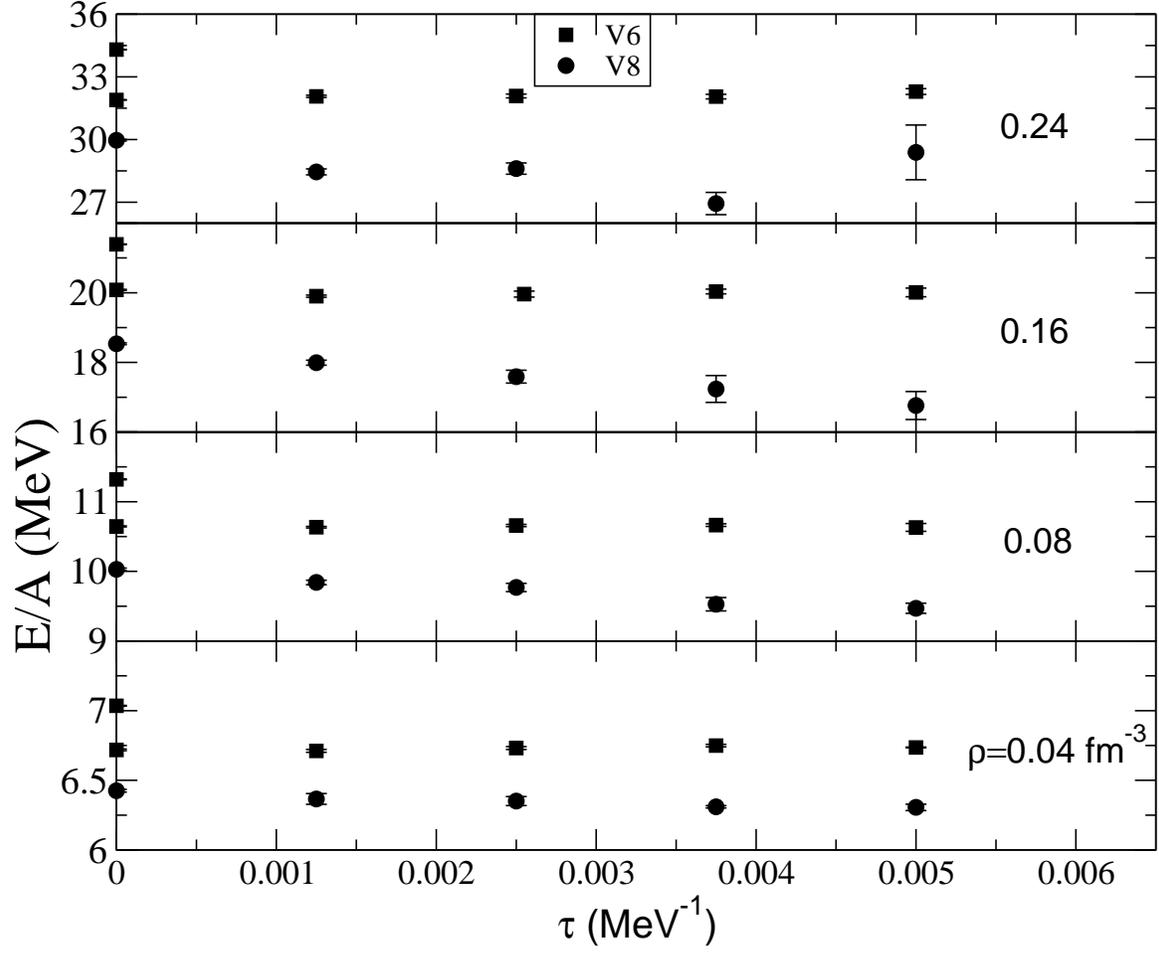}
\caption{Energy vs. imaginary time $\tau$ after CP propagation, 
at various densities.  VMC results (upper squares) 
for $\vsp$ and GFMC-CP results (lower squares and 
dots) for $\vsp$ and $\vep$ are shown at $\tau = 0$,
and unconstrained GFMC results are shown for various $\tau > 0$.}
\label{fig:evstau}
\end{figure}

\begin{figure}
\includegraphics[width=6in,angle=270]{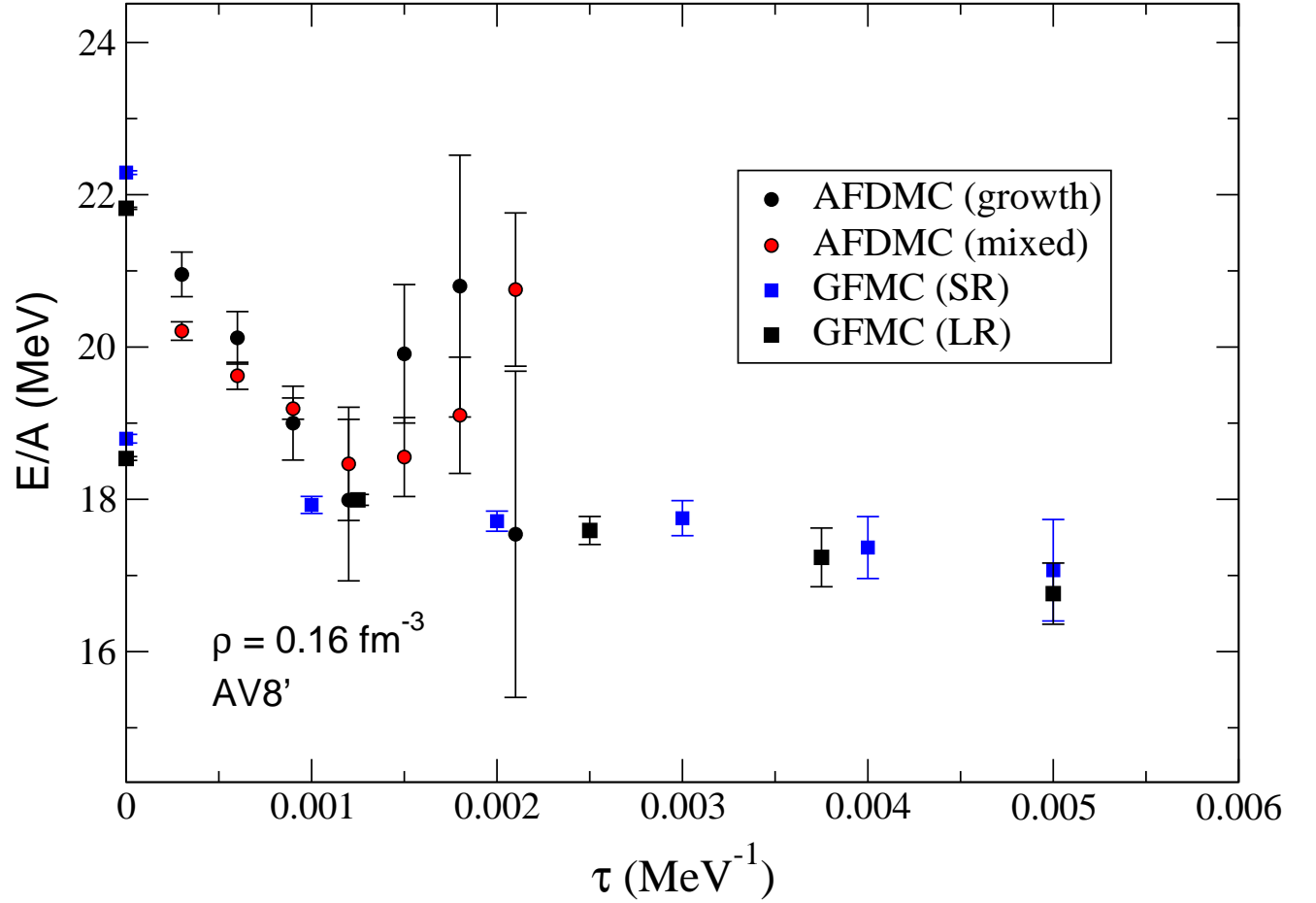}
\caption{Energy vs. imaginary time $\tau$ for different calculations 
using $\vep$ interaction at $\rho = $ 0.16 fm$^{-3}$.
Two different estimates (growth and mixed) are shown for the AFDMC
calculation along with the results of two different GFMC calculations, 
using short-range (SR) and long-range (LR) correlations in the trial 
wave function. } 
\label{fig:gfmcafmc}
\end{figure}

\newpage
\begin{figure}
\includegraphics[width=6in]{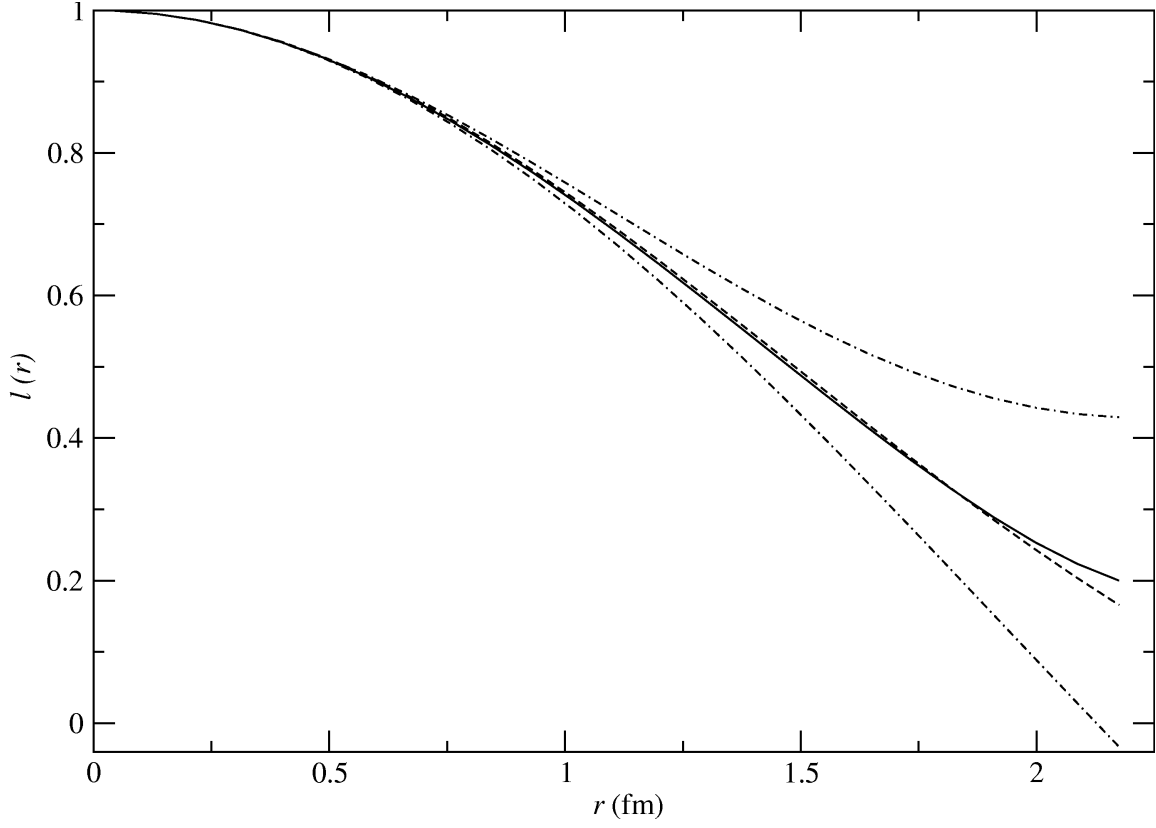}
\caption{The Slater function $\ell_{PB}({\bf r})$  for 14 neutrons
in PB.  The top and bottom dash-dot lines show the $\ell_{PB}({\bf r})$ 
with ${\bf r}$ parallel to the 0,0,1 box side and to the 1,1,1 diagonal 
respectively.  The middle full line shows the average $\sqrt{\overline{
\ell_{PB}^2({\bf r})}}$, and the dashed line shows the $\ell(r)$ in the UG.} 
\label{fig:boxl}
\end{figure}

\newpage
\begin{figure}
\includegraphics[width=5in]{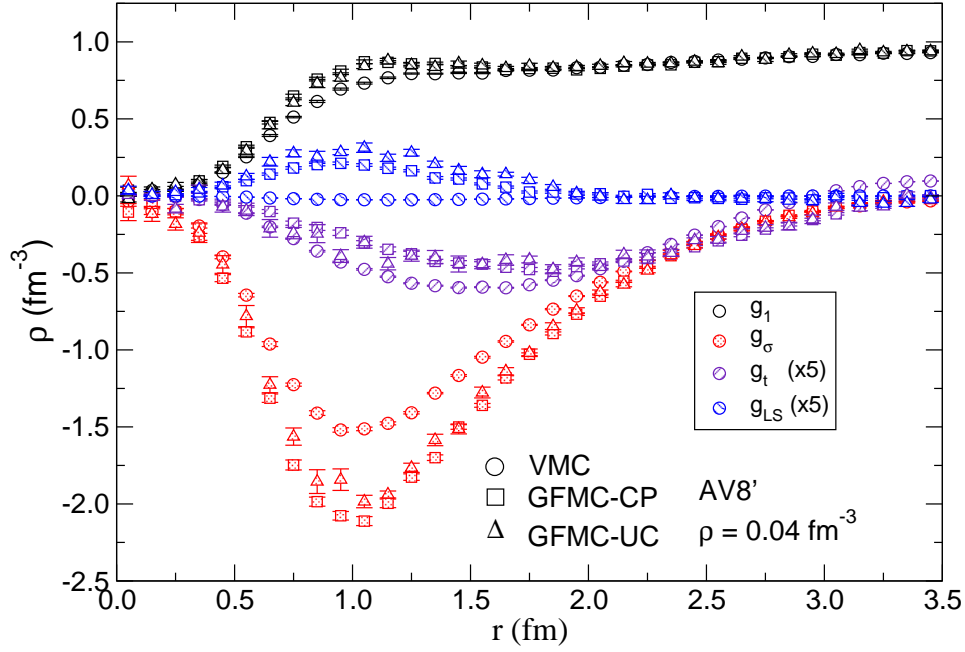}
\caption{Pair distribution functions for the $\vep$ interaction
at rho = 0.04 fm$^{-3}$. 
The data sets from top to bottom correspond to 
central, L$\cdot$S, tensor and $\boldsigma_i \cdot \boldsigma_j$ 
pair distribution functions. Each set contains circles, squares and 
triangles showing the VMC, GFMC-CP and GFMC-UC results. The L$\cdot$S 
and the tensor distribution functions are scaled up by a factor of five. }  
\label{fig:pairdisv8-.04}
\end{figure}

\begin{figure}
\includegraphics[width=5in]{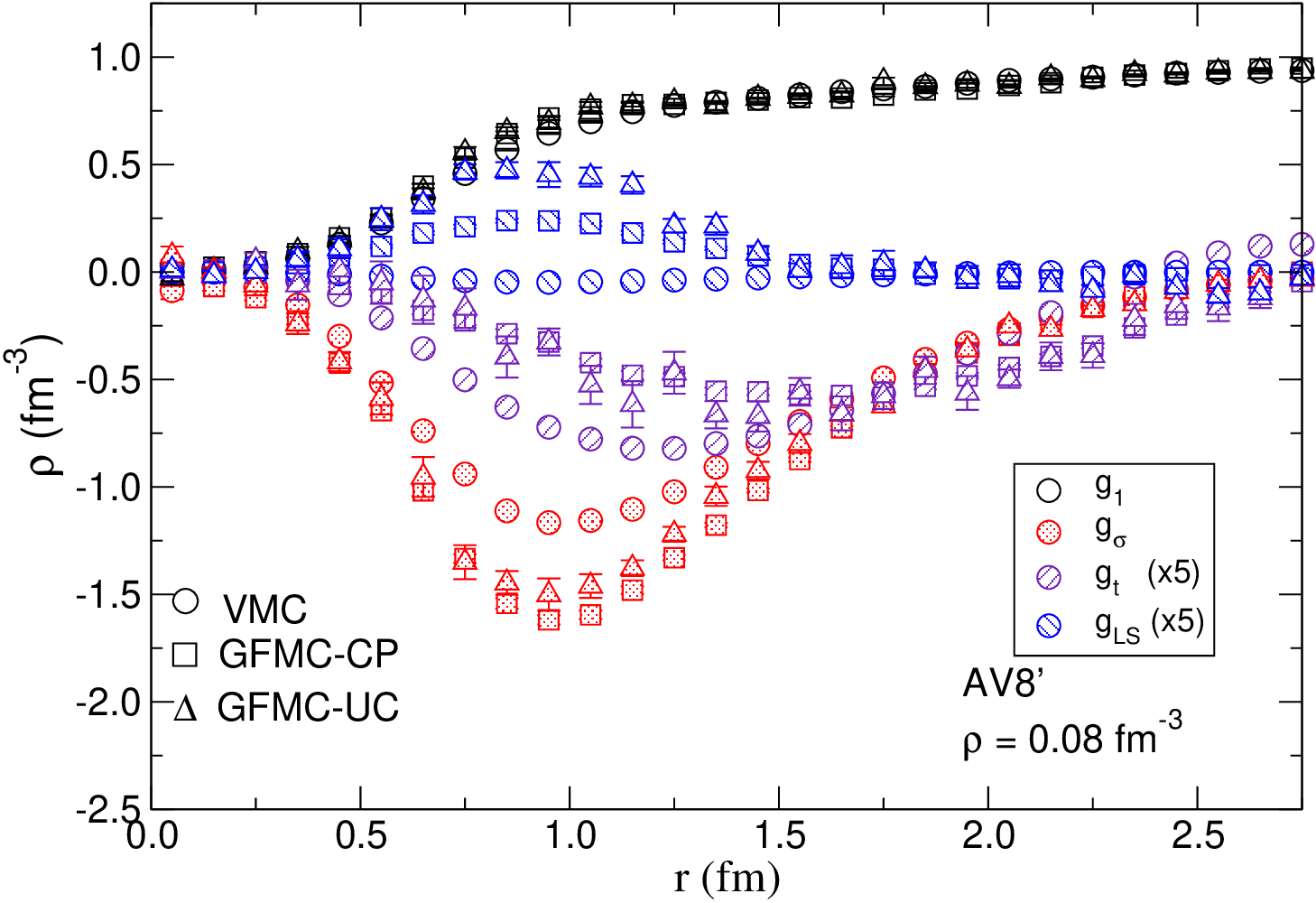}
\caption{Pair distribution functions for the $\vep$ interaction
at rho = 0.08 fm$^{-3}$, as in Figure \ref{fig:pairdisv8-.04}.}
\label{fig:pairdisv8-.08}
\end{figure}

\begin{figure}
\includegraphics[width=5in]{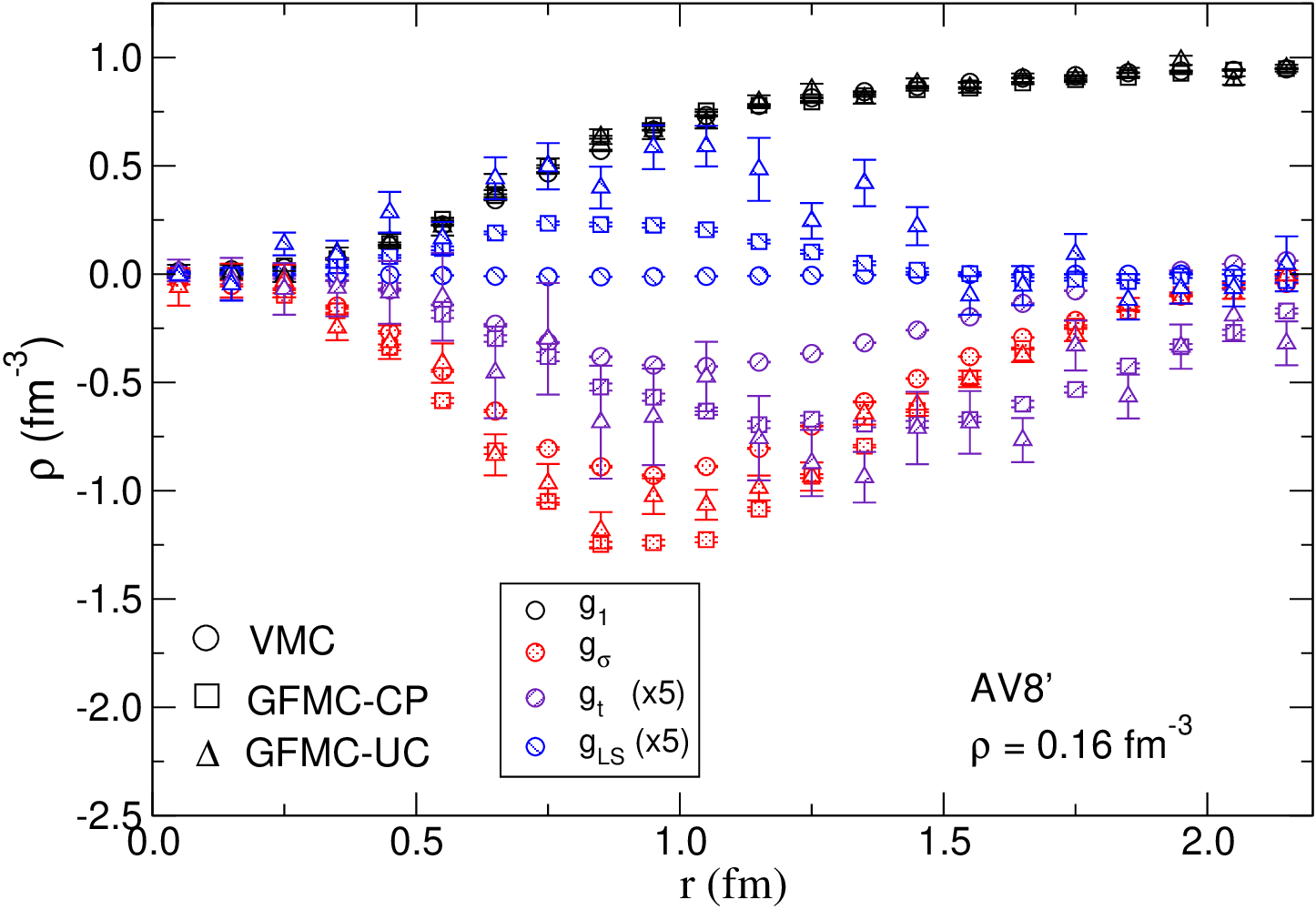}
\caption{Pair distribution functions for the $\vep$ interaction
at rho = 0.16 fm$^{-3}$, as in Figure \ref{fig:pairdisv8-.04}.}
\label{fig:pairdisv8-.16}
\end{figure}

\begin{figure}
\includegraphics[width=5in]{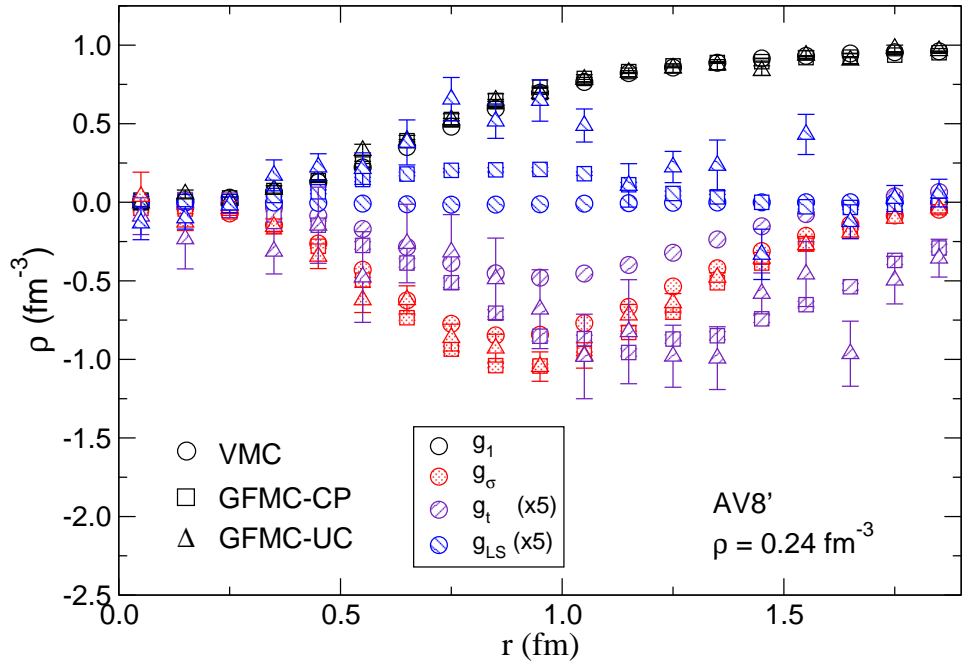}
\caption{Pair distribution functions for the $\vep$ interaction
at rho = 0.24 fm$^{-3}$, as in Figure \ref{fig:pairdisv8-.04}.}
\label{fig:pairdisv8-.24}
\end{figure}

\begin{figure}
\includegraphics[width=4in]{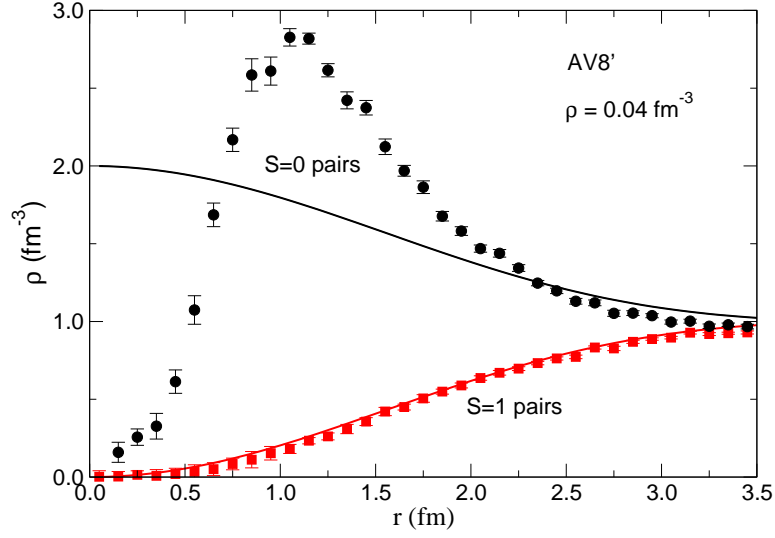}
\caption{Pair distribution functions for spin 0 and spin 1 pairs 
at $\rho = 0.04$ fm$^{-3}$; results of  
unconstrained GFMC calculations are compared to distributions in 
noninteracting FG shown by solid lines.}
\label{fig:spinzero}
\end{figure}

\begin{figure}
\includegraphics[width=4in,angle=0]{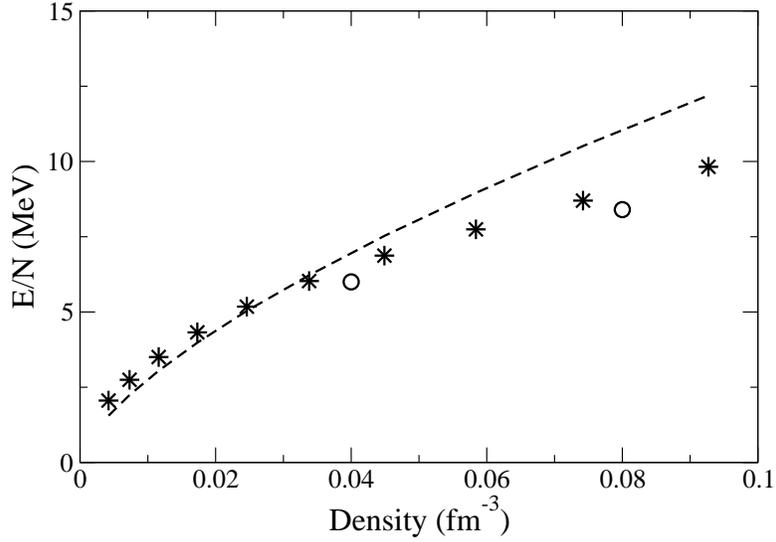}
\caption{The energy of neutron matter at low densities.  The stars give
the 
results of Ref.\cite{FP81} and the circles of the present calculation.  The dashed
line 
shows $0.5~E_{FG}$. } 
\label{fig:ldnm}
\end{figure}

\begin{figure}
\includegraphics[width=4in,angle=0]{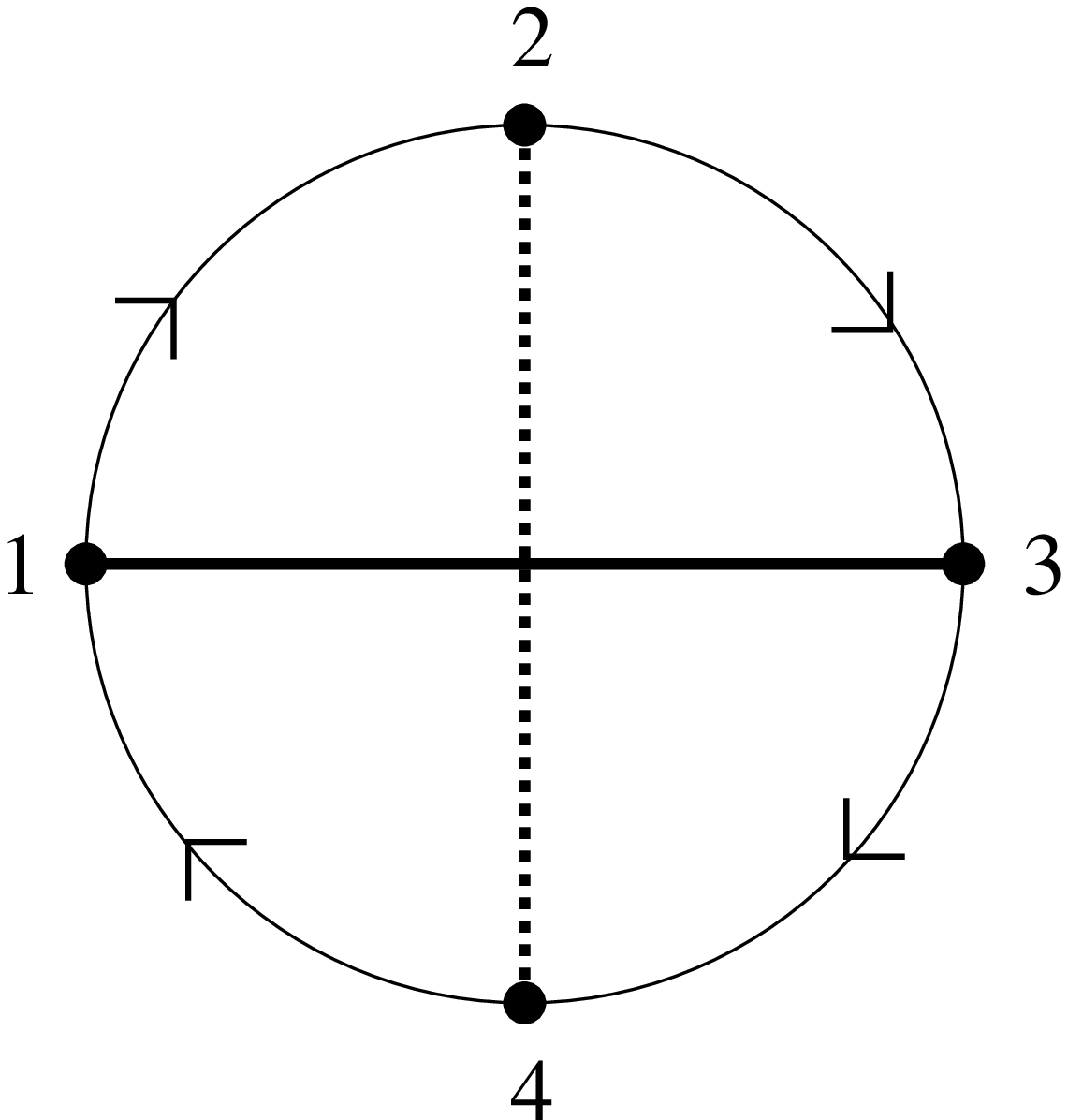}
\caption{The lowest order elementary 4-body circular exchange diagram.  The dashed line 
shows $F_{24}^2-1$, the thick solid line denotes the interaction link, and the thin 
lines with direction show the exchange pattern.} 
\label{fig:ele4cc}
\end{figure}


\begin{thebibliography}{99}


\bibitem{PR95} C. J. Pethick and D. G. Ravenhall, 
	Annu. Rev. Nucl. Part. Sci. {\bf 45}, 429 (1995). 
\bibitem{HP00} H. Heiselberg and V. R. Pandharipande, 
	Annu. Rev. Nucl. Part. Sci. {\bf 50}, 481 (2000). 
\bibitem{WNX} W. Nazarewicz in {\em The RIA White paper},
        www.nscl.msu.edu/future/ria/process/ whitepapers/durham2000.pdf (2000). 
\bibitem{Engvik97} L. Engvik, M. Hjorth-Jensen, R. Machleidt, H.
M\"{u}ther  
	and A. Polls, Nucl. Phys. {\bf A627}, 85 (1997); L. Engvik, {\em The 
	nuclear equation of state,} Thesis, Univ. of Oslo (1999). 
\bibitem{BBB97} M. Baldo, I. Bombaci and G. F. Burgio, Astron.
Astrophys.
        {\bf 328}, 274 (1997).
\bibitem{WFF88} R. B. Wiringa, V. Fiks and A. Fabrocini,
        Phys. Rev. C {\bf 38}, 1010 (1988).
\bibitem{APR98} A. Akmal, V. R. Pandharipande and D. G. Ravenhall,
        Phys. Rev. C {\bf 58}, 1804 (1998).
\bibitem{SBGL98} H. Q. Song, M. Baldo, G. Giansiracusa and U. Lombardo,
        Phys. Rev. Lett. {\bf 81}, 1584 (1998).
\bibitem{MPR02} J. Morales, V. R. Pandharipande and D. G. Ravenhall, 
	Phys. Rev. C {\bf 66}, 054308 (2002). 
\bibitem{WPCP00} R. B. Wiringa, S. C. Pieper, J. Carlson and V. R.
Pandharipande,
        Phys. Rev. C {\bf 62}, 014001 (2000).
\bibitem{PWV02} S. C. Pieper, K. Verga and R. B. Wiringa, 
        Phys. Rev. C {\bf 66}, 044310 (2002).
\bibitem{PPWC01} S. C. Pieper, V. R. Pandharipande, R. B. Wiringa and J.
Carlson, Phys. Rev. C {\bf 64}, 014001 (2001).
\bibitem{SF99}  K. E. Schmidt and S. Fantoni, 
	Phys. Lett. {\bf B446}, 99 (1999). 
\bibitem{FSS01} S. Fantoni, A. Sarsa and K.E. Schmidt,
        Phys Rev Lett {\bf 87}, 181101 (2001).
\bibitem{Bertsch} G. F. Bertsch, {\em Challenge Problem in Many-Body 
	Physics} www.phys.washington.edu /~mbx/george (1998)
\bibitem{ERS97} J. R. Engelbrecht and  M. Randeria and  C.A.R. {S{\'a} de Melo},
      Phys. Rev. B {\bf 55}, 15153 (1997).
\bibitem{dfg} J. Carlson, V. R. Pandharipande, K. E. Schmidt and S-Y Chang, 
private comunication (2002). 
\bibitem{PPCPW97} B. S. Pudliner, V. R. Pandharipande, J. Carlson, S. C.
Pieper 
	and R. B. Wiringa, Phys. Rev. C {\bf 56}, 1720 (1997). 
\bibitem{lheqmc} D. M. Ceperley, Rev. Mod. Phys. 67, 279 (1995).
\bibitem{egasqmc} G. Ortiz and P. Ballone, Phys. Rev. B {\bf 50},
(1994).
\bibitem{fixednode} J. B. Anderson, J. Chem. Phys. {\bf 63},
1499 (1975). 
\bibitem{constrainedpath} Shiwei Zhang, J. Carlson, and J. E.
Gubernatis,
Phys. Rev. B{\bf 55} , 7464 (1997); J. Carlson, J. E. Gubernatis, G.
Ortiz, and
Shiwei Zhang, Phys. Rev. B{\bf 59}, 12788 (1999).
\bibitem{PW79} V. R. Pandharipande and R. B. Wiringa,  
        Rev. Mod. Phys. {\bf 51}, 821 (1979).
\bibitem{PBFHNC} S. Fantoni and K. E. Schmidt, Nucl. Phys. A{\bf 690},
456, 2001.
\bibitem{BCLL92} M. Baldo, J. Cugnon, A. Lejeune and U. Lombardo, 
	Nucl. Phys. {\bf A536}, 349 (1992). 
\bibitem{AP97} A. Akmal and V. R. Pandharipande, Phys. Rev. C {\bf 56}, 
	2261 (1997). 
\bibitem{FWbook} A. L. Fetter and J. D. Walecka, {\em Quantum Theory of
Many Particle Systems}, McGraw-Hill (1971).
\bibitem{baker} G. A. Baker, Phys. Rev. C {\bf 60}, 054311 (1999). 
\bibitem{hh} H. Heiselberg, Phys. Rev. A {\bf 63}, 043606 (2001). 
\bibitem{FP81} B. Friedman and V. R. Pandharipande, Nucl. Phys. {\bf
A361}, 501 (1981). 
\bibitem{PRlect} V. R. Pandharipande and D. G. Ravenhall, in {\em 
Proc. NATO Advanced Research Workshop on Nuclear Matter and Heavy Ion 
Collisions, Les Houches}, Ed. M. Soyeur et.al., pp 103 Plenum (1989). 
\bibitem{zab81} J. G. Zabolitzky, Adv. Nucl. Phys. {\bf 12}, 1 (1981). 
\bibitem{krot79} E. Krotscheck, Nucl. Phys. {\bf A317}, 149 (1979). 
\bibitem{npa473} R. Schiavilla, D. S. Lewart, V. R. Pandharipande, S. C. 
Pieper, R. B. Wiringa and S. Fantoni, Nucl. Phys. {\bf A473}, 267 (1987). 

\end{thebibliography}
\end{document}